\documentclass{elsart}
\usepackage{epsfig}
\usepackage{enumerate}

\begin{document}
\begin{frontmatter}

\title{A three-state kinetic agent-based model to analyze tax evasion dynamics}

\author{Nuno Crokidakis}
\thanks{nuno@if.uff.br}

\address{
Instituto de F\'{\i}sica, \hspace{1mm} Universidade Federal Fluminense \\
Av. Litor\^anea s/n, \hspace{1mm} 24210-340 \hspace{1mm} Niter\'oi - RJ, \hspace{1mm} Brazil}

\maketitle

\begin{abstract}
\noindent
In this work we study the problem of tax evasion on a fully-connected population. For this purpose, we consider that the agents may be in three different states, namely honest tax payers, tax evaders and undecided, that are individuals in an intermediate class among honests and evaders. Every individual can change his/her state following a kinetic exchange opinion dynamics, where the agents interact by pairs with competitive negative (with probability $q$) and positive (with probability $1-q$) couplings, representing agreement/disagreement between pairs of agents. In addition, we consider the punishment rules of the Zaklan econophysics model, for which there is a probability $p_{a}$ of an audit each agent is subject to in every period and a length of time $k$ detected tax evaders remain honest. Our results suggest that below the critical point $q_{c}=1/4$ of the opinion dynamics the compliance is high, and the punishment rules have a small effect in the population. On the other hand, for $q>q_{c}$ the tax evasion can be considerably reduced by the enforcement mechanism. We also discuss the impact of the presence of the undecided agents in the evolution of the system.

\end{abstract}
\end{frontmatter}


\section{Introduction}

\qquad In the recent years, the statistical physics techniques have been successfully applied in the description of socioeconomic phenomena. Among the studied problems we can cite opinion dynamics, language evolution, biological aging, dynamics of stock markets, earthquakes and many others \cite{galam_book,sen_book,econ_book,pmco_book}. These interdisciplinary topics are usually treated by means of computer simulations of agent-based models, which allow us to understand the emergence of collective phenomena in those systems.

A challenging interdisciplinary subject is tax evasion dynamics, which is an interesting practical topic to be studied because tax evasion remains to be a major predicament facing governments \cite{bloom,prinz,andreoni}. Models of tax evasion were firstly studied by economists \cite{gachter,frey,follmer,slemrod,davis}, and more recently physicists became also interested in the subject \cite{zaklan,lima1,lima2,lima3,llacer,seibold} (for recent reviews, see \cite{bloom,prinz}). Experimental evidence provided by Gachter suggests that tax payers tend to condition their decision regarding whether to pay taxes or not on the tax evasion decision of the members of their group \cite{gachter}. In addition, Frey and Torgler also provide empirical evidence on the relevance of conditional cooperation for tax morale \cite{frey}. Based on these ideas, Zaklan \textit{et al.} recently proposed a model that has been attracted attention \cite{zaklan}. In the so-called Zaklan model, the dynamics of tax payers and tax evaders is analyzed by means of the two-dimensional Ising model at a given temperature $T$. In this model, each agent $i$ may be in one of two possible states, namely $s_{i}=+1$ (honest) or $s_{i}=-1$ (cheater or tax evader). A transition $s_{i} \to -s_{i}$ (or a spin flip) is controlled by the ``social temperature'' $T$ and also depends on the nearest neighbors' states of the agent (or spin) at site $i$. Thus, for low temperatures few spin flips occur and for high temperatures many spin flips occur. In other words, tax evaders have the greatest influence to turn honest citizens into tax evaders if they constitute a majority in the respective neighborhood. In addition, some punishment rules are applied: there is a probability $p_{a}$ of an audit each agent is subject to in every period and a length of time $k$ detected tax evaders remain honest \cite{zaklan}. In another work, the dynamics of the model was also controlled by another two-state model, namely the majority-vote model with noise \cite{maj_vot}, where the noise $q$ plays the role of the temperature. In this case, similar results were found \cite{lima3}, suggesting that the results of the Zaklan model are robust.

In this work we study the tax evasion dynamics by means of a three-state agent-based model. The agents interact by pairs considering kinetic exchanges of their states, in a way that the pairwise couplings may be positive of negative. In addition, we apply the punishment rules of the Zaklan econophysics model. Our results suggest that above the critical point of the opinion dynamics the tax evasion can be considerably reduced by the enforcement rules. On the other hand, below the critical point the compliance is high, and the punishment rules have a small impact on the evasion.

This work is organized as follows. In Section 2 we present the microscopic rules that define the model, and the numerical results are discussed in Section 3. Finally, our conclusions are presented in Section 4.


\section{Model}

\qquad Our model is based on a kinetic exchange opinion model \cite{biswas}. A population of $N$ agents is defined on a fully-connected graph, i.e.,  each agent can interact  with all others. In opposition to what occurs in the Zaklan model \cite{zaklan}, for which the dynamics is governed by the Ising model (i.e., a two-state model), in our model each individual $i$ ($i=1,2,...,N$) carries one of three possible states or attitudes at a given time step $t$, represented by $s_{i}(t)=+1,-1$ or $0$. The dynamic rules are defined following the opinion model of Ref. \cite{biswas}. Each social interaction occurs between two given agents $i$ and $j$, and we considered that $j$ will influence $i$. First, this pair of agents $(i,j)$ is randomly chosen. Then, the state of the agent $i$ in the next time step $t+1$ will be updated according to
\begin{equation}\label{eq1}
s_{i}(t+1) = {\rm sgn}\left[ s_{i}(t) + \mu_{ij}\,s_{j}(t)  \right]\,,
\end{equation}
where the sign function is defined such that ${\rm sgn}(0)=0$ and the interaction strenghts $\{\mu_{ij}\}$ are quenched random variables given by the discrete bimodal probability distribution
\begin{equation}\label{eq2}
F(\mu_{ij}) = q\,\delta(\mu_{ij}+1) + (1-q)\,\delta(\mu_{ij}-1) ~.
\end{equation}

Notice that we considered that each agent can in principle interact with all other agents, i.e. there is no specific underlying topology for the structure of the interaction network. So the model can be viewed as an infinite dimension (or ``mean field'') Zaklan model. This is an almost realistic situation thanks to the modern social and communication networks. 

First, let us elaborate upon the nature of the above-mentioned three states. The state $s_{i}=+1$ represents a honest tax payer, i.e., an individual 100$\%$ convinced of his/her honesty, who does not consider evasion. He/she is either habitually compliant or he/she is a recent evader who has become honest as a result of enforcement efforts or social norms. On the other hand, the state $s_{i}=-1$ represents a cheater, i.e, an individual who is an evading tax payer. Whether a tax payer continues to evade depends on both enforcement and the effect of social interactions.

Those two classes correspond to the $\pm 1$ states of the standard Zaklan model \cite{zaklan}. In addition, we have considered a third state, $s_{i}=0$, which can be interpreted as an \textit{undecided} individual. However, notice that the above rules of the opinion dynamics [Eqs. (\ref{eq1}) and (\ref{eq2})] impose that for an agent to shift from state $s=+1$ to $s=-1$ or vice-versa it must to pass by the intermediate undecided state $s=0$ \cite{biswas}. Thus, an agent that is currently at the state $s=0$ was a honest tax payer ($s=+1$) or a tax evader ($s=-1$) before. In the first case, the individual is a honest tax payer and, due to social interactions, he/she becomes a tax payer who is dissatisfied with the tax system, perhaps as a result of seeing others evading without being punished. He/she is not actively evading, but he/she might if the perceived benefits of doing so exceed the perceived costs. For this group, evasion is an option. On the other hand, the second possibility is that the agent is a tax evader and, due to social interactions, he/she stops temporarily the evasion because he/she wondered whether it is worth to evade. This agent is fickle and is not 100$\%$ convinced of his/her honesty, and thus he/she can become a honest tax payer ($s=+1$) or he/she can come back to the tax evader state ($s=-1$), depending on the next interactions with his/her social contacts. The above discussion will become more clear in the following, when we will discuss the interpretation of the competitive interactions $\mu_{ij}$.

The pairwise couplings $\mu_{ij}$ in Eq. (\ref{eq2}) may be either negative (with probability $q$) or positive (with probability $1-q$), such that $q$ represents the fraction of negative couplings \cite{biswas,celia}. The above process given by Eqs. (\ref{eq1}) and (\ref{eq2}) is repeated $N$ times, which defines one time step in the dynamics. In addition to such basic dynamics of the model, after the $N$ interactions we have considered a policy makers' tax enforcement mechanism consisting of two components, a probability $p_{a}$ of an audit each person is subject to in every period and a length of time $k$ detected tax evaders remain honest, as considered in the Zaklan model \cite{zaklan}. In other words, after the application of the above-mentioned kinetic exchange opinion dynamics, we have considered that each tax evader will be caught by an audit with probability $p_{a}$. In this case, the individual must remain honest for a given number $k$ of time steps. As mentioned before, the above rules of the opinion dynamics impose that for an agent to shift from state $s_{i}=+1$ to $s_{i}=-1$ or vice-versa it must to pass by the intermediate undecided state $s_{i}=0$. However, this kind of hierarchy is partially broken when we apply the enforcement rules. Indeed, if a tax evader (state $s_{i}=-1$) is caught by an audit, he/she changes directly to the honest state $s_{i}=+1$ and remains in this state at least during the next $k$ time steps. 

Following ref. \cite{biswas}, the dynamics defined by the kinetic exchange model can be interpreted as follows. If $\mu_{ij}$ is positive (with probability $1-q$), there is a kind of agreement between the agents $i$ and $j$. In this case, if $s_{i}(t)=0$ the agent $i$ is undecided and does not know what is the best choice (be honest or be evader), and thus he/she follows the decision of $j$, i.e., the state of $i$ is updated to $s_{i}(t+1)=s_{j}(t)$. If $s_{i}(t)=s_{j}(t)$ nothing occurs, and if $s_{i}(t)=-s_{j}(t)$ the agent $i$ will become undecided and change to state $s_{i}(t+1)=0$ in the next time step $t+1$, and he/she can become evader or a honest tax payer in a next interaction. On the other hand, if $\mu_{ij}$ is negative (with probability $q$), there is a kind of disagreement or mutual disliking between the agents $i$ and $j$. In this case, if $s_{i}(t)=s_{j}(t)$ the agent $i$ will become undecided and change to $s_{i}(t+1)=0$ and and if $s_{i}(t)=-s_{j}(t)$ the agent $i$ keeps his/her decision. Finally, if $s_{i}(t)=0$, the agent $i$ does not know what is the best choice (be honest or be evader) and due to the mentioned disliking his state is updated to $s_{i}(t+1)=-s_{j}(t)$.

\begin{figure}[t]
\begin{center}
\vspace{6mm}
\includegraphics[width=0.5\textwidth,angle=0]{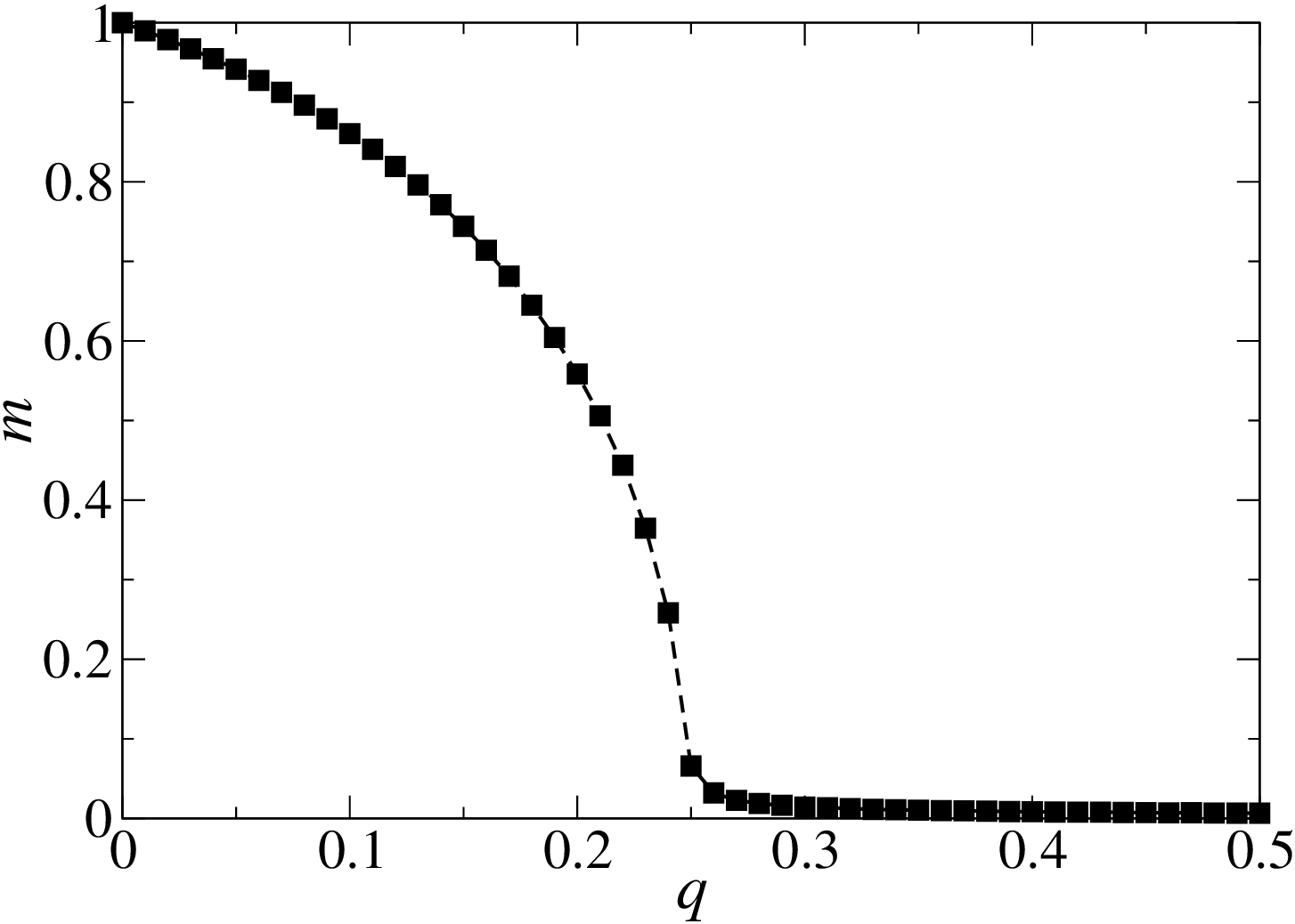}
\end{center}
\caption{Magnetization per spin $m$ of the kinetic exchange opinion model of Ref. \cite{biswas} as a function of the fraction $q$ of negative interactions (i.e., where no punishment rules were considered). The system undergoes an order-disorder phase transition at $q_{c}=1/4$, with a paramagnetic disordered phase defined by the coexistence of the three states $s=+1, -1$ and $0$ with equal fractions ($1/3$ for each one). The population size is $N=10^{4}$, the squares are numerical results averaged over $100$ independent simulations and the dashed line is just a guide to the eyes.}
\label{fig1}
\end{figure}

As discussed in \cite{biswas}, the standard opinion dynamics defined by Eqs. (\ref{eq1}) and (\ref{eq2}), i.e., where no punishment rules were considered, undergoes a nonequilibrium phase transition at a critical fraction $q_{c}=1/4$. For $q<q_{c}$ we have a symmetry of the extreme opinions $s=\pm 1$, i.e., one of the extreme opinions $+1$ or $-1$ dominates the system, with consensus states occurring only for $q=0$, i.e., in the absence of negative interactions. On the other hand, for $q\geq q_{c}$ the system is in a  disordered ``paramagnetic'' phase characterized by the coexistence of the three opinions, with the fraction of each opinion being $1/3$. This picture can be clearly seen in Fig. \ref{fig1}, where we exhibit the order parameter of the system, i.e., 
\begin{equation} \label{eq3}
m = \left\langle \frac{1}{N}\left|\sum_{i=1}^{N} s_{i}\right|\right\rangle ~, 
\end{equation}
where $\langle\, ...\, \rangle$ denotes a disorder or configurational average taken at steady states. The Eq. (\ref{eq3}) defines the ``magnetization per spin'' of the system, and the behavior of $m$ as a function of the fraction $q$ of negative interactions for a population of size $N=10^{4}$ agents is shown in Fig. \ref{fig1}. 

Thus, the next step is to apply the enforcement rules of the Zaklan model \cite{zaklan} to the model of Ref. \cite{biswas}. The numerical results will be discussed in the next section.


\section{Numerical Results}

\qquad We applied the enforcement rules of the Zaklan model \cite{zaklan} to the opinion dynamics model of Ref. \cite{biswas}. As usually occurs in Zaklan-like models \cite{zaklan,lima1,lima2,lima3}, we have considered that initially all agents are honest, i.e., we have $s_{i}(t=0)=+1$ for all individuals $i$. In this case, we broke the above-mentioned symmetry of the stationary states of the ordered phase: for $q<q_{c}$ the majority of agents will be in the honest state $s_{i}=+1$. All the following results are for a population of size $N=10^{4}$.

\begin{figure}[t]
\begin{center}
\vspace{6mm}
\includegraphics[width=0.48\textwidth,angle=0]{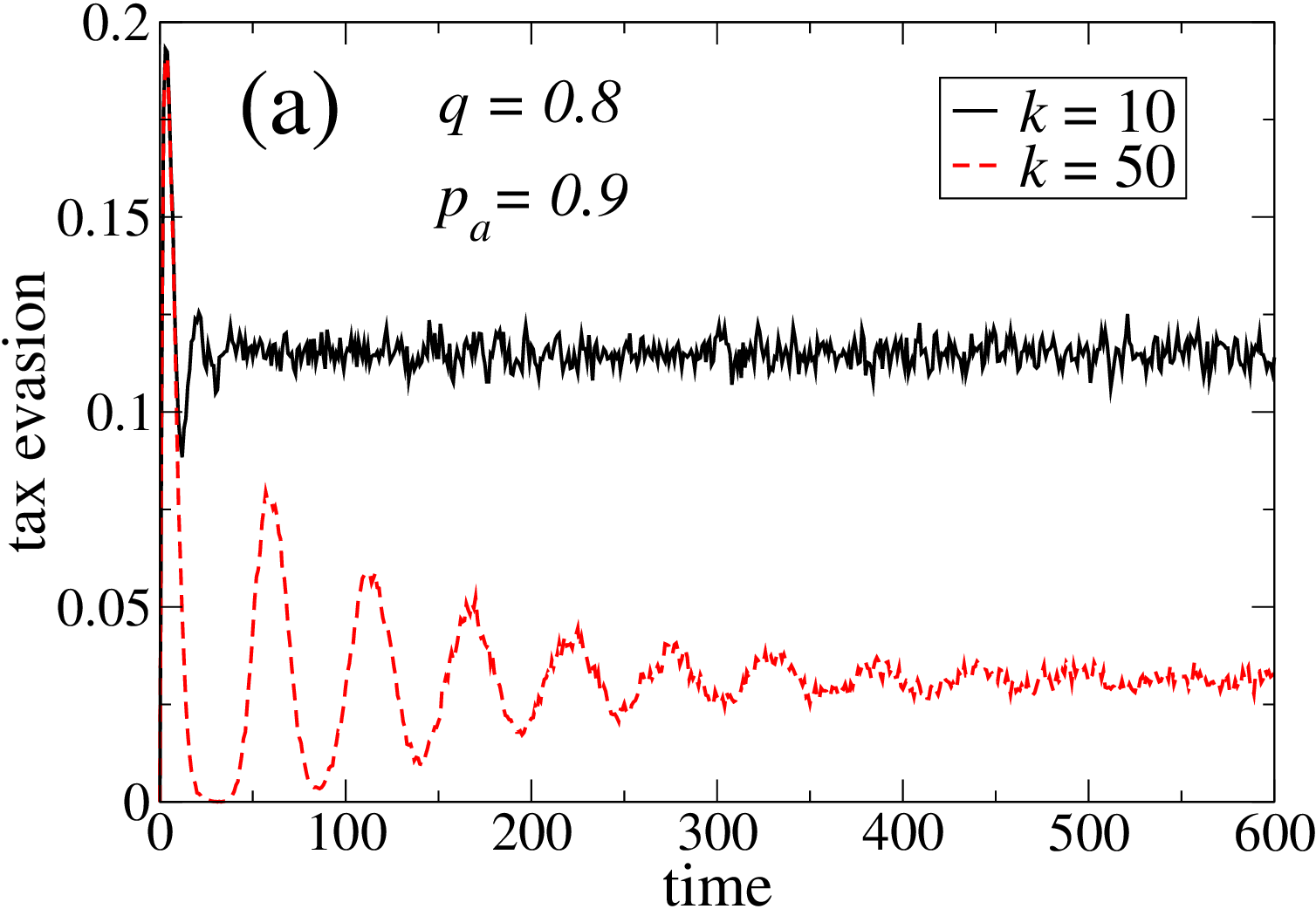}
\hspace{0.3cm}
\includegraphics[width=0.47\textwidth,angle=0]{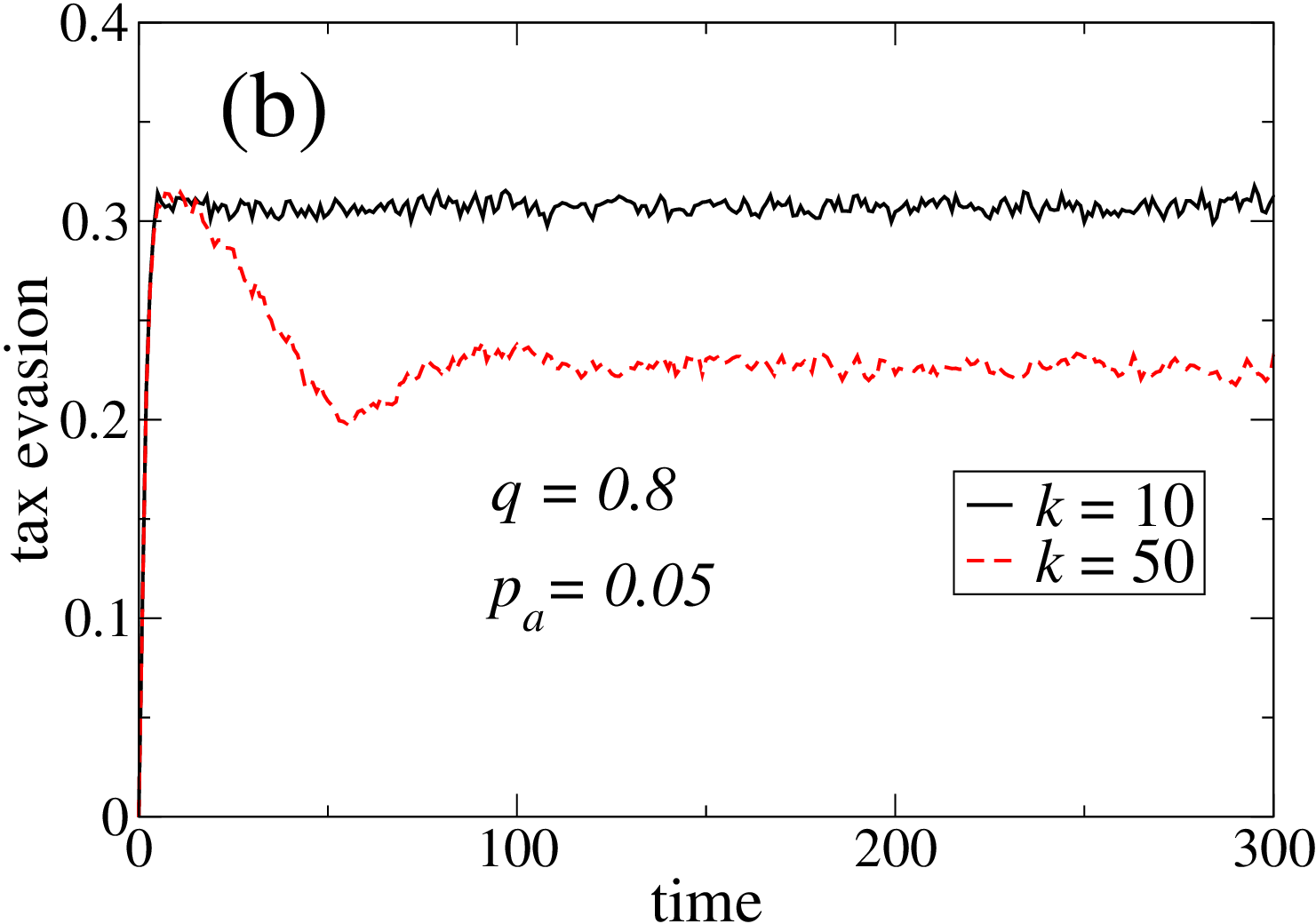}
\\
\vspace{1.0cm}
\includegraphics[width=0.48\textwidth,angle=0]{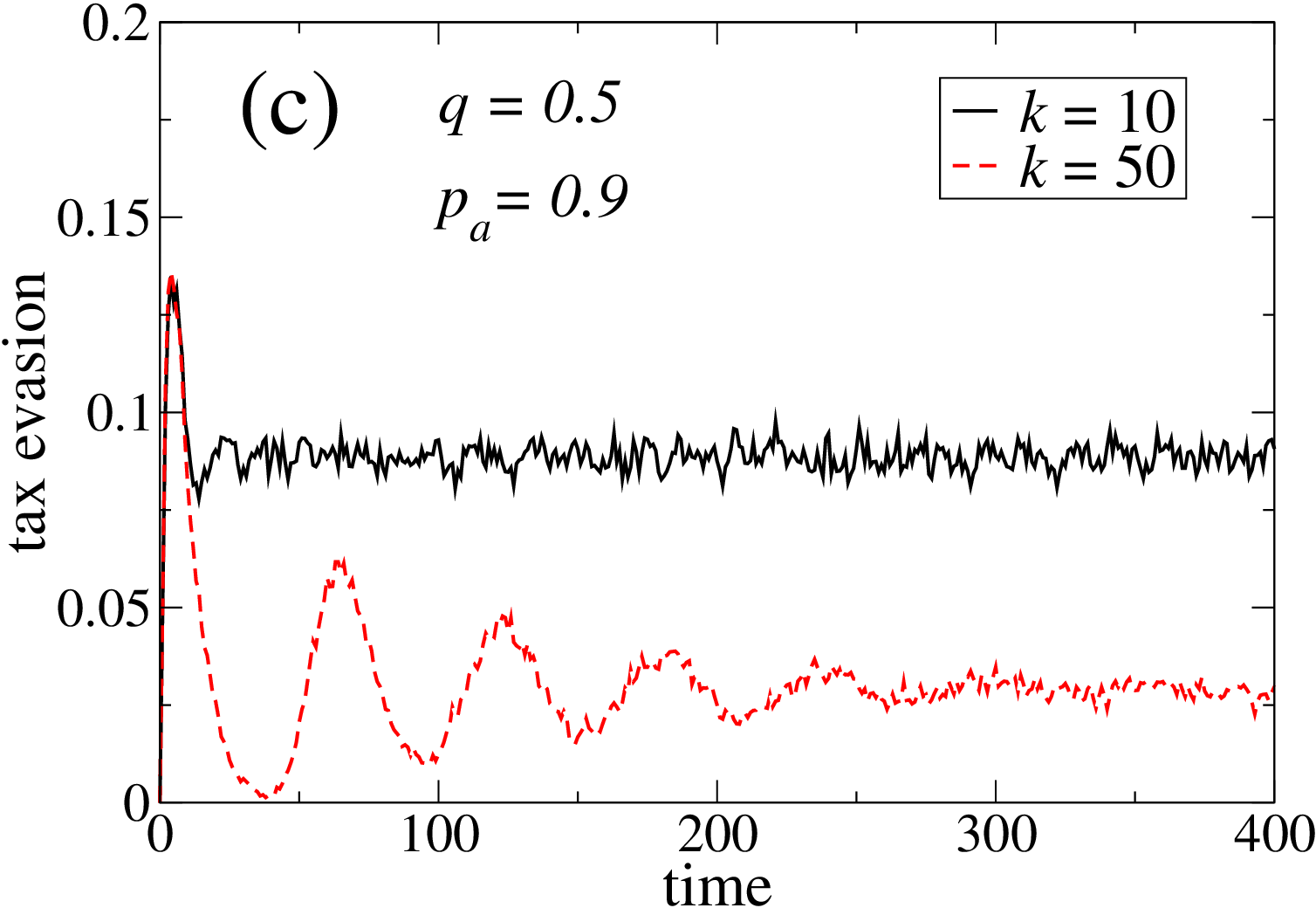}
\hspace{0.3cm}
\includegraphics[width=0.47\textwidth,angle=0]{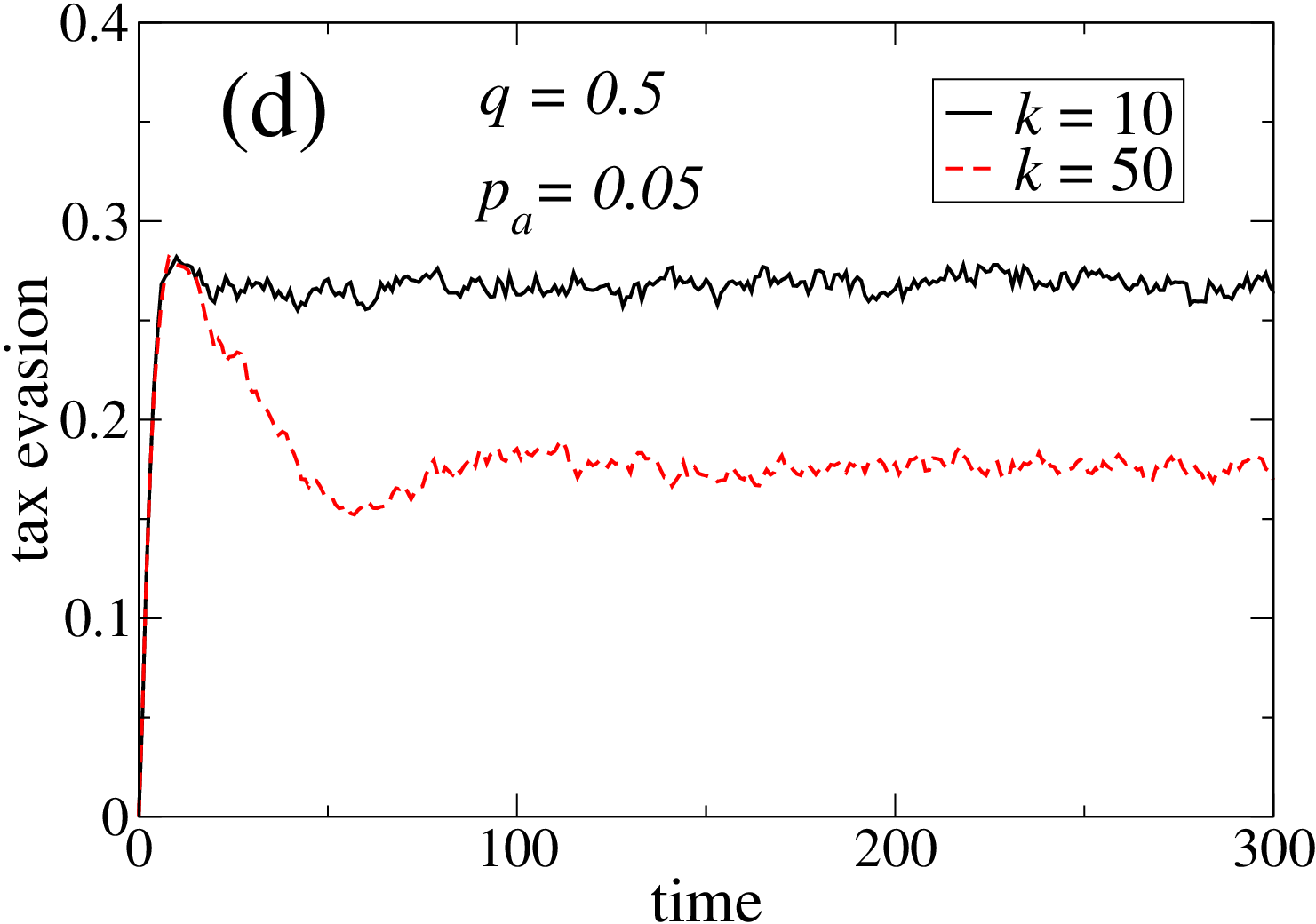}
\end{center}
\caption{(Color online) Time evolution of the tax evasion for different values of the number $k$ of periods that a detected tax evader must remain honest for and two distinct audit probabilities, namely $p_{a}=0.9$ (left side) and $p_{a}=0.05$ (right side). The results are for $q=0.8$ [(a) and (b)] and $q=0.5$ [(c) and (d)]. Each curve is a single realization of the dynamics for a population of size $N=10^{4}$.}
\label{fig2}
\end{figure}

Following the previous studies of the Zaklan model \cite{zaklan,lima1,lima2,lima3}, one can start analyzing the time evolution of the tax evasion, i.e., the fraction of tax evaders ($s=-1$) in the population. In Fig. \ref{fig2} we exhibit the tax evasion as a function of time for two distinct values of $q>q_{c}$, namely $q=0.8$ [(a) and (b)] and $q=0.5$ [(c) and (d)]. In this case, as $q>q_{c}$, the kinetic exchange dynamics defined by Eqs. (\ref{eq1}) and (\ref{eq2}) leads the system to a disordered state with an equal fraction of each state. In other words, considering only the opinion dynamics, the stationary fraction of evaders should be $1/3\approx 33\%$. Thus, one can see from Fig. \ref{fig2} that if the audits are efficient ($p_{a}=90\%$) the tax evasion can be considerably reduced to $\approx 10\%$ for $k=10$ and for $\approx 3\%$ for $k=50$. In these cases, we observe similar fluctuations of the tax evasion as the ones reported in Zaklan models defined in regular lattice and networks \cite{zaklan,lima1,lima2,lima3}. For the cases where we consider a realistic value of the efficient audits ($p_{a}=5\%$) the punishment is effective only if the penalty duration is high ($k=50$). In this case, the tax evasion can be reduced for values around $20\%$. Notice that when we decrease the value of $q$ the fraction of tax evaders decreases, as one can see in Fig. \ref{fig1}. It can be understood as follows. As the opinion dynamics ``coexists'' in the system with the punishment rules, the system does not achieve in our model the steady states with equal fractions of the three opinions $+1,-1$ and $0$ for $q>q_{c}$. For high values of $q$, there are many negative couplings $\mu_{ij}$ in the population, which allows many transitions $s_{i}=+1 \to s_{i}=0$ and then $s_{i}=0 \to s_{i}=-1$. So, it is expected that for high $q$ the fraction of opinions $-1$ is greater than in the cases of lower values of $q$ (of course, we are talking about the disordered phase of the kinetic exchange opinions dynamics). 

As discussed above, when we decrease the value of the parameter $q$, the fraction of tax evaders decreases (for $q>q_{c}$). In this case, the compliance increases when we apply the same punishments for a population with lower values of $q$. Furthermore, one can conclude that for a population with a large fraction of negative interactions, i.e., with a high disagreement among the individuals, the compliance is low in the system, and thus it is necessary a strong enforcement by the public policies in order to control the tax evasion, which is a realistic feature of the model.

\begin{figure}[t]
\begin{center}
\vspace{6mm}
\includegraphics[width=0.48\textwidth,angle=0]{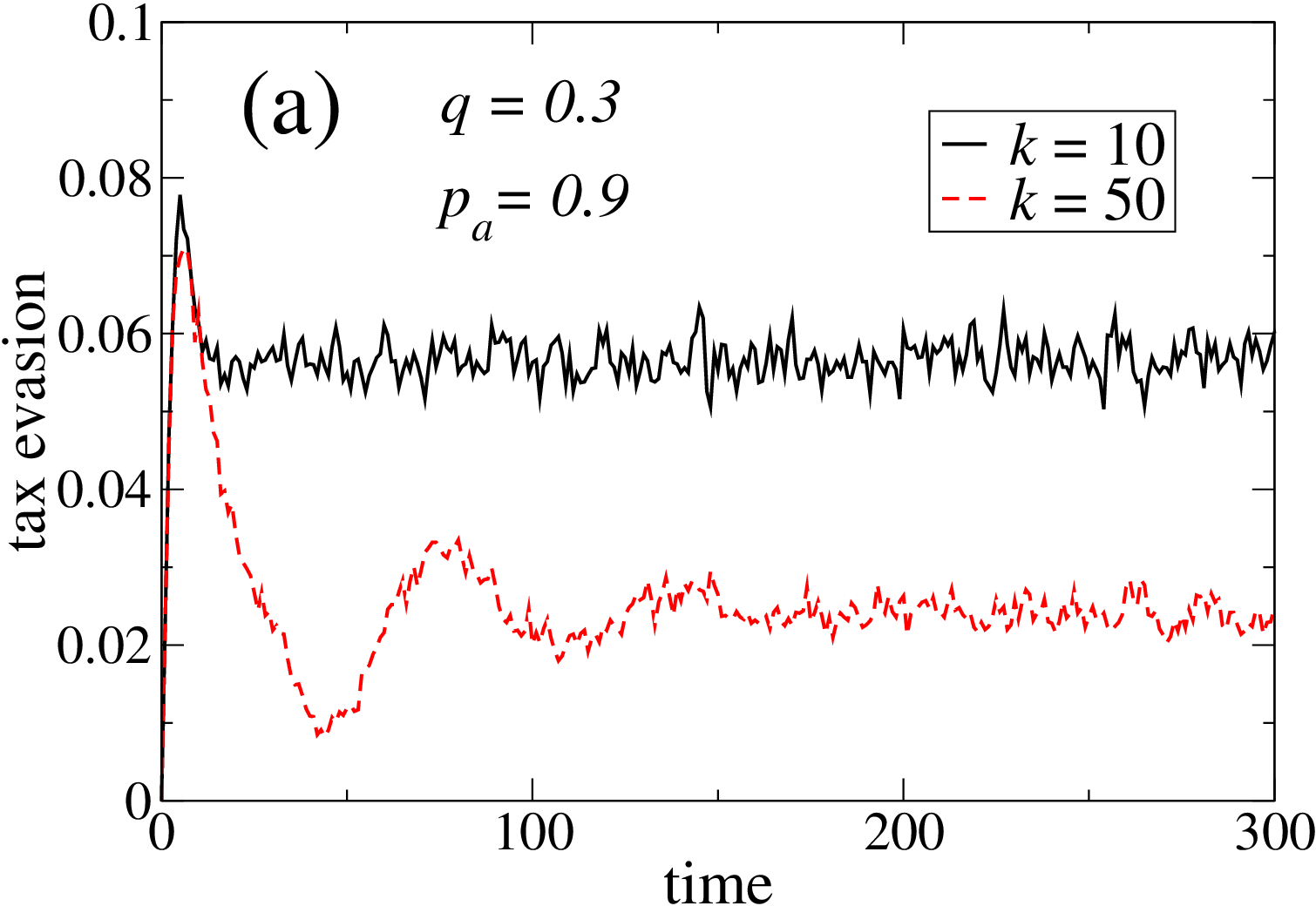}
\hspace{0.3cm}
\includegraphics[width=0.48\textwidth,angle=0]{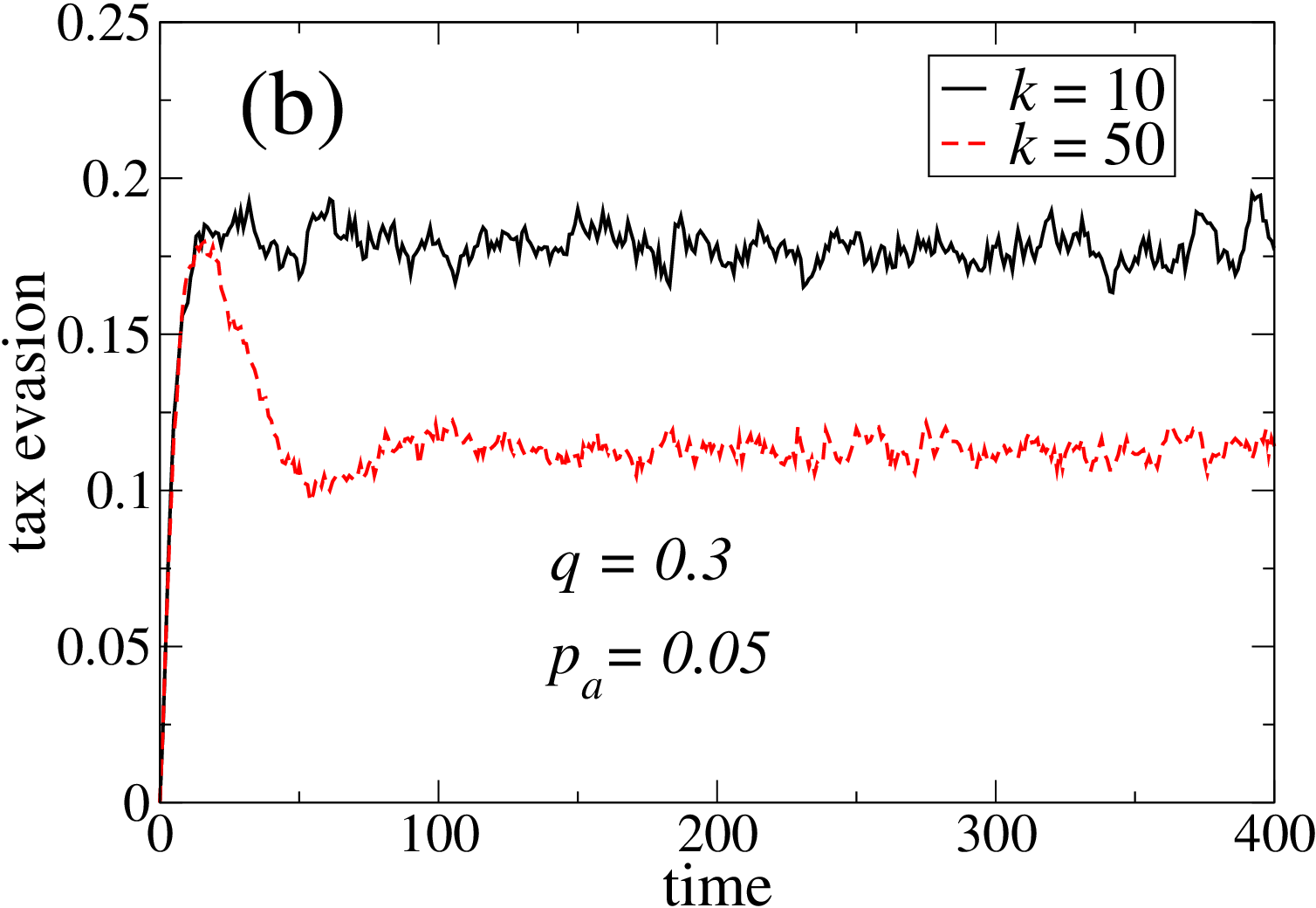}
\end{center}
\caption{(Color online) Time evolution of the tax evasion for $q=0.3$, different values of the number $k$ of periods that a detected tax evader must remain honest for and two distinct audit probabilities, namely $p_{a}=0.9$ (a) and $p_{a}=0.05$ (b). Each curve is a single realization of the dynamics for a population of size $N=10^{4}$.}
\label{fig3}
\end{figure}

\begin{figure}[t]
\begin{center}
\vspace{5mm}
\includegraphics[width=0.48\textwidth,angle=0]{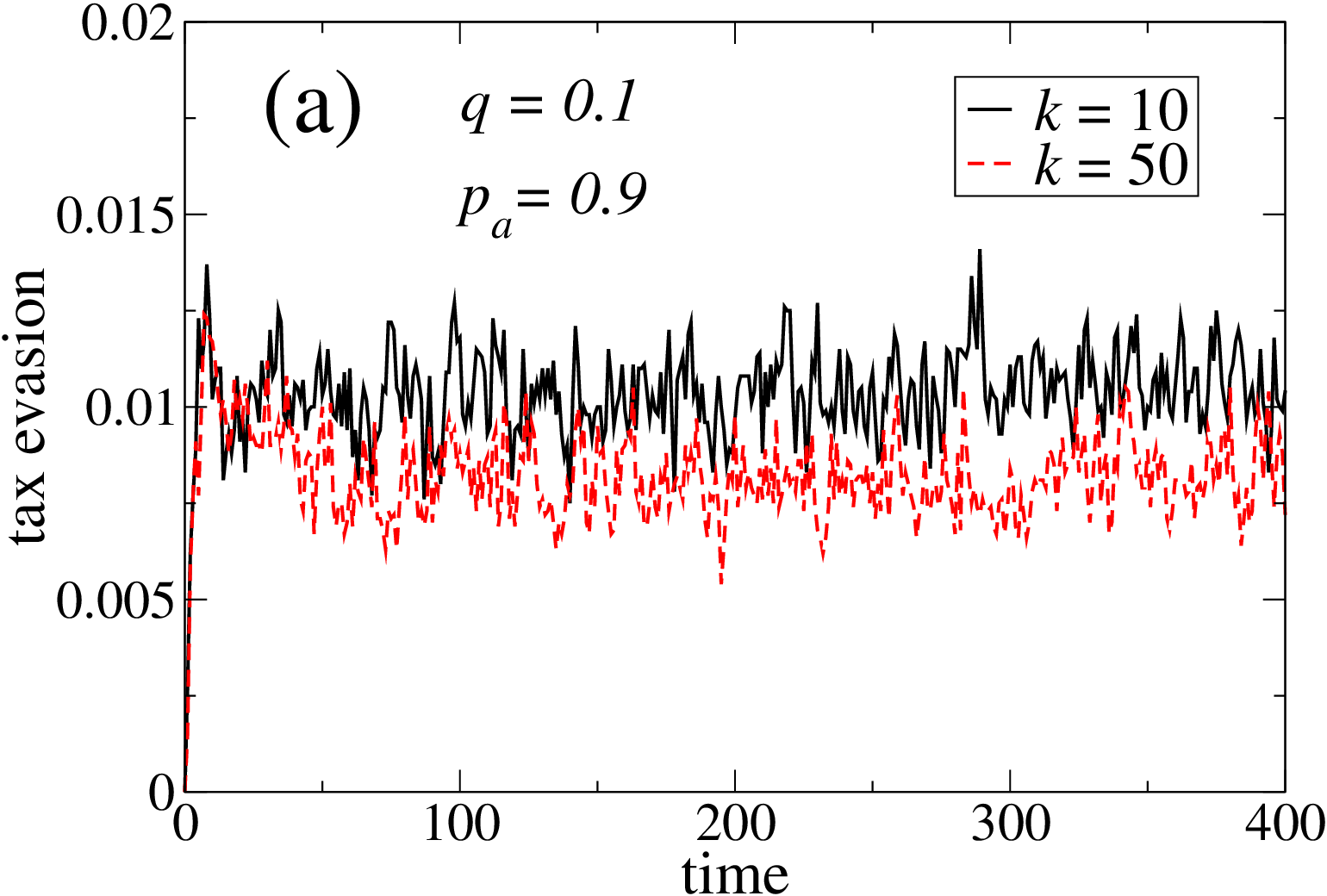}
\hspace{0.3cm}
\includegraphics[width=0.48\textwidth,angle=0]{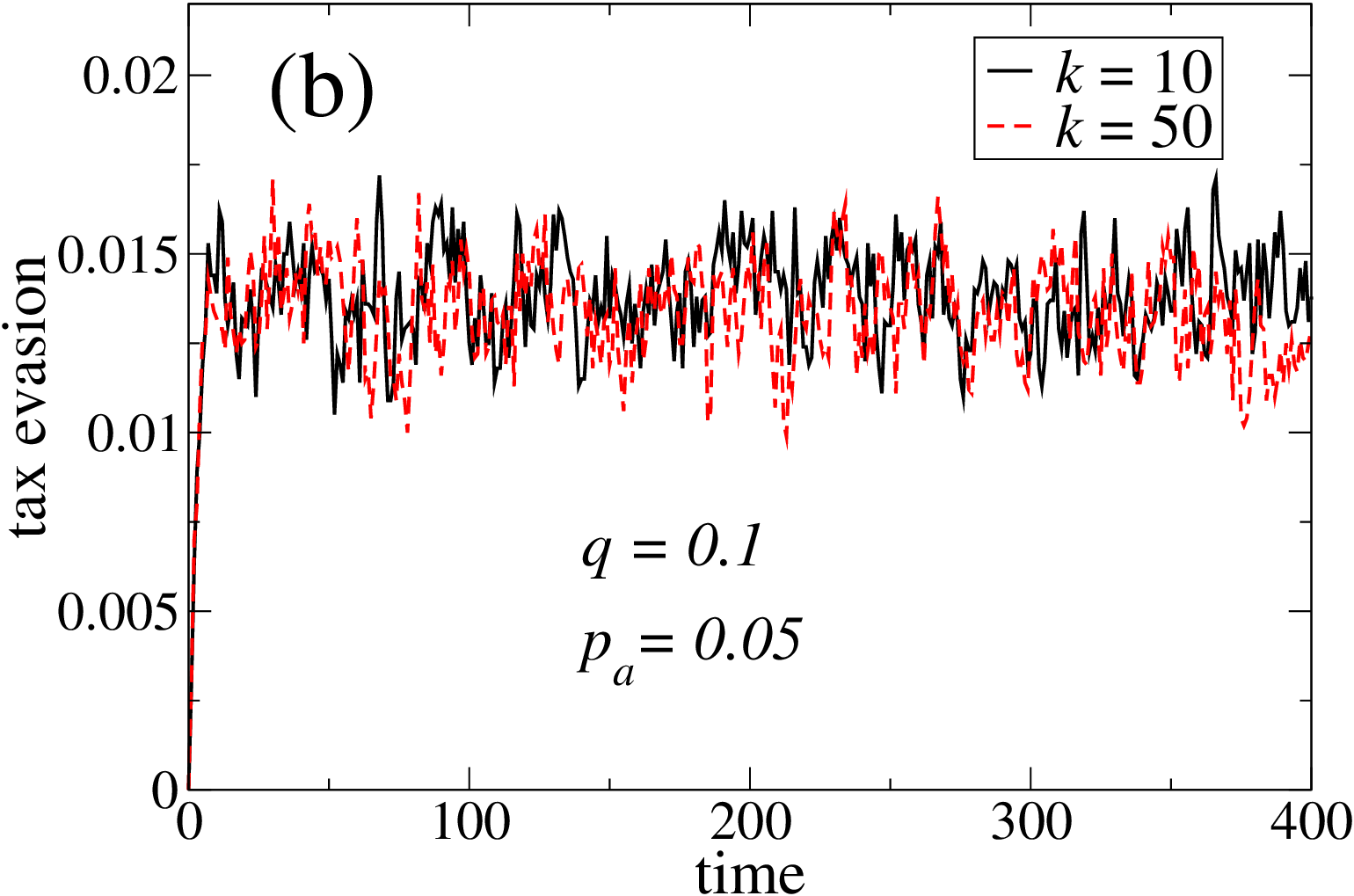}
\end{center}
\caption{(Color online) Time evolution of the tax evasion for $q=0.1$, different values of the number $k$ of periods that a detected tax evader must remain honest for and two distinct audit probabilities, namely $p_{a}=0.9$ (a) and $p_{a}=0.05$ (b). Each curve is a single realization of the dynamics for a population of size $N=10^{4}$.}
\label{fig4}
\end{figure}

In Fig. \ref{fig3} we exhibit results for $q=0.3$, i.e., another value of $q>q_{c}$, but now the system is near the critical point $q_{c}=1/4$. In this case, one can see that for a high audit probability like $p_{a}=90\%$ the tax evasion can be extremely reduced in a short run [see Fig. \ref{fig3} (a)], even if only a small period like $k=10$ each individual is compelled to remain honest. Notice that even for $p_{a}=5\%$ the application of severe punishments as $k=50$ can lead the evasion to low levels like $10\%$, as one can see in Fig. \ref{fig3} (b).

It is shown in Fig. \ref{fig4} the time series of the tax evasion for $q=0.1$. In this case, as $q<q_{c}$, the kinetic exchange dynamics defined by Eqs. (\ref{eq1}) and (\ref{eq2}) leads the system to an ordered state with the majority of agents in the honest ($s=+1$) state, with a small fraction of tax evaders and undecided. For $N=10^{4}$ (see Fig. \ref{fig1}), the fraction of $s=-1$ individuals is $\approx 1.5\%$. Thus, as one can see in Fig. \ref{fig4}, the consideration of the punishment rules togheter with the basic dynamics has a small effect on the tax evasion in the system, since the tendency of the agents is to be in state $s=+1$ and thus there are few tax evaders to be caught by the audits. For $p_{a}=5\%$ the stationary fraction of evaders is similar for small or large $k$, and for $p_{a}=90\%$ the tax evasion decreases to $\approx 1\%$, with a small difference between the cases $k=10$ and $k=50$. Thus, for a population with a small fraction of negative interactions, i.e., with low disagreement (or high agreement) among the individuals, the compliance is high in the system, and it is not necessary a strong control by the public policies.

\begin{figure}[t]
\begin{center}
\vspace{5mm}
\includegraphics[width=0.48\textwidth,angle=0]{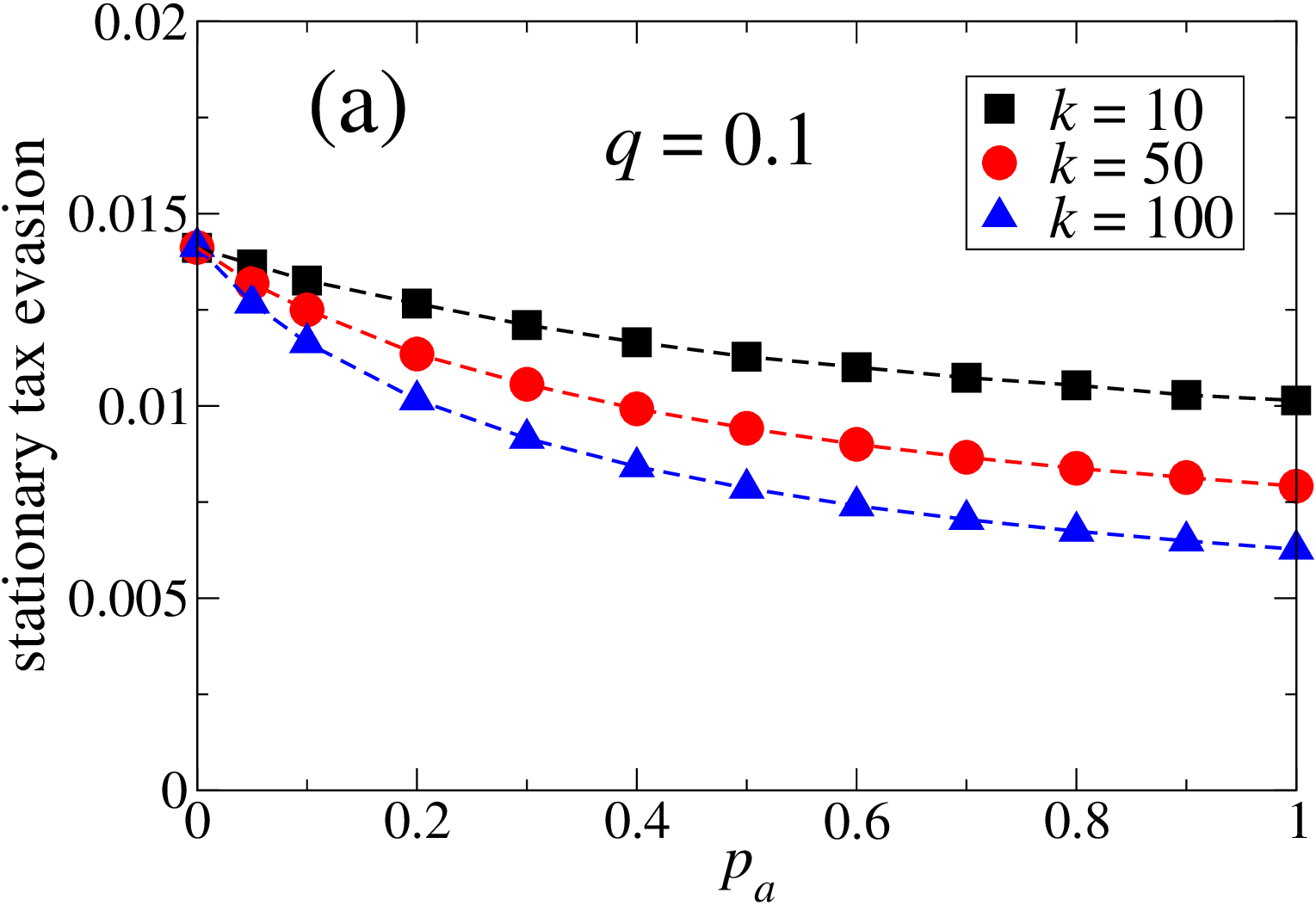}
\hspace{0.3cm}
\includegraphics[width=0.47\textwidth,angle=0]{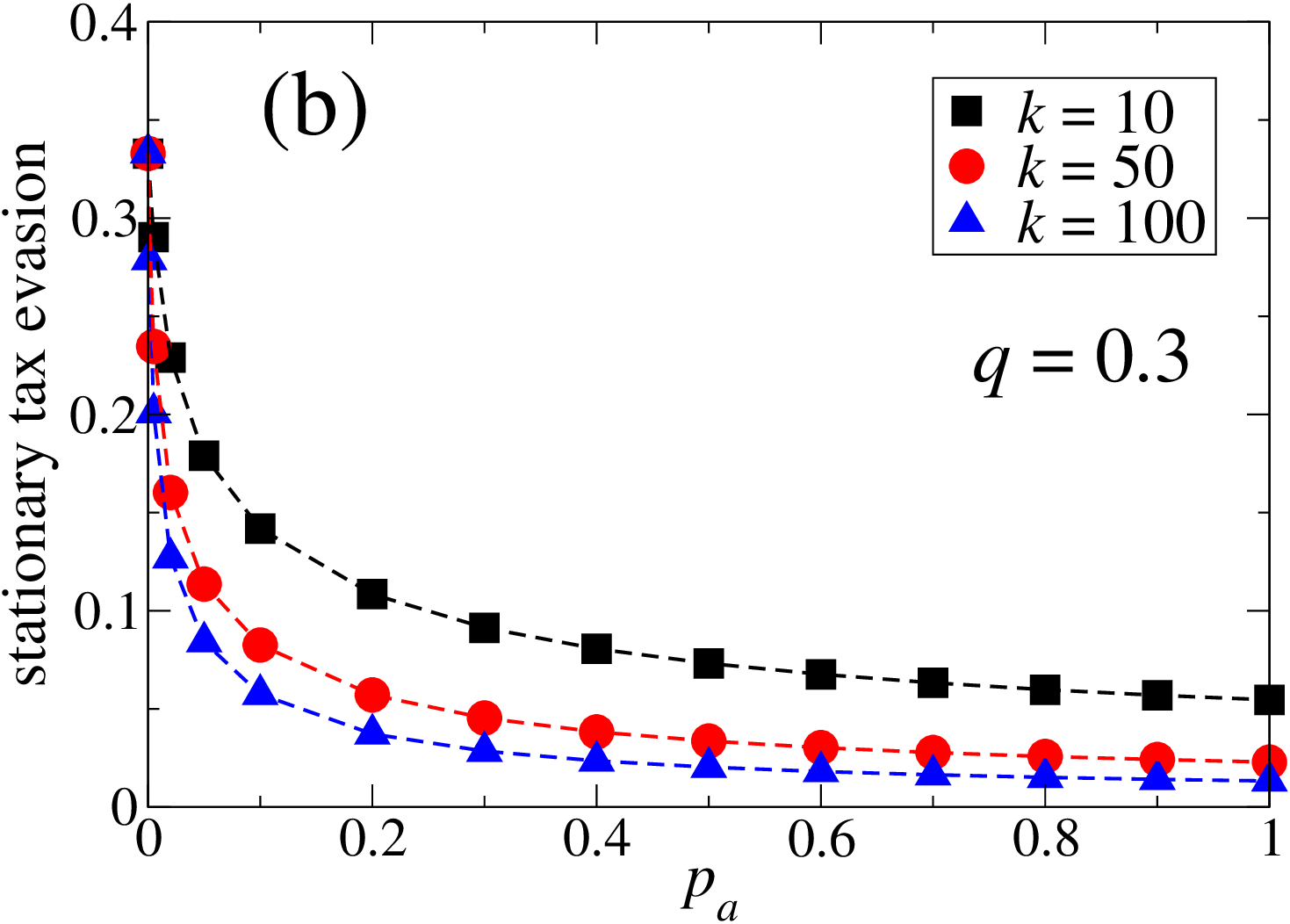}
\\
\vspace{1.0cm}
\includegraphics[width=0.47\textwidth,angle=0]{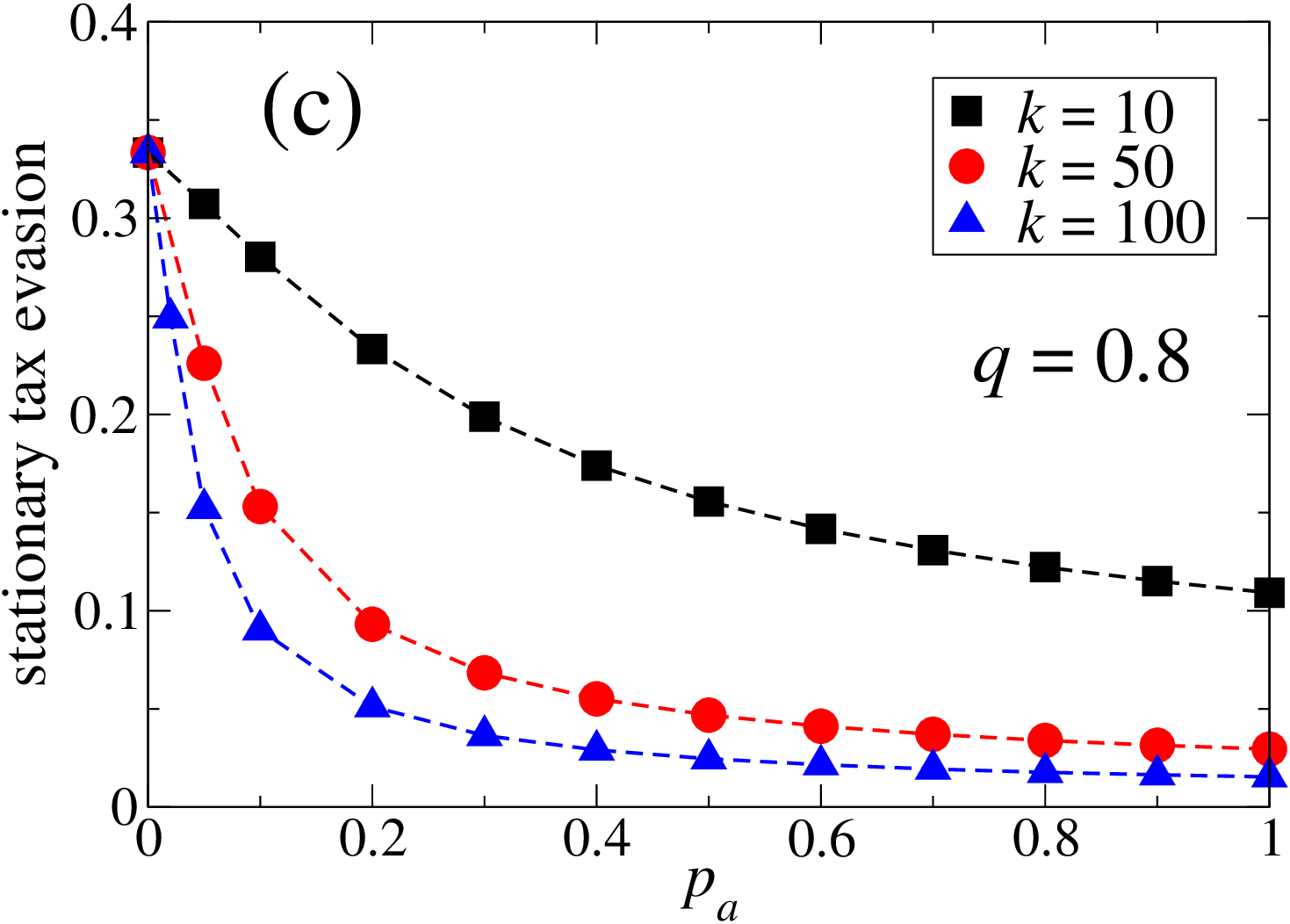}
\hspace{0.3cm}
\includegraphics[width=0.47\textwidth,angle=0]{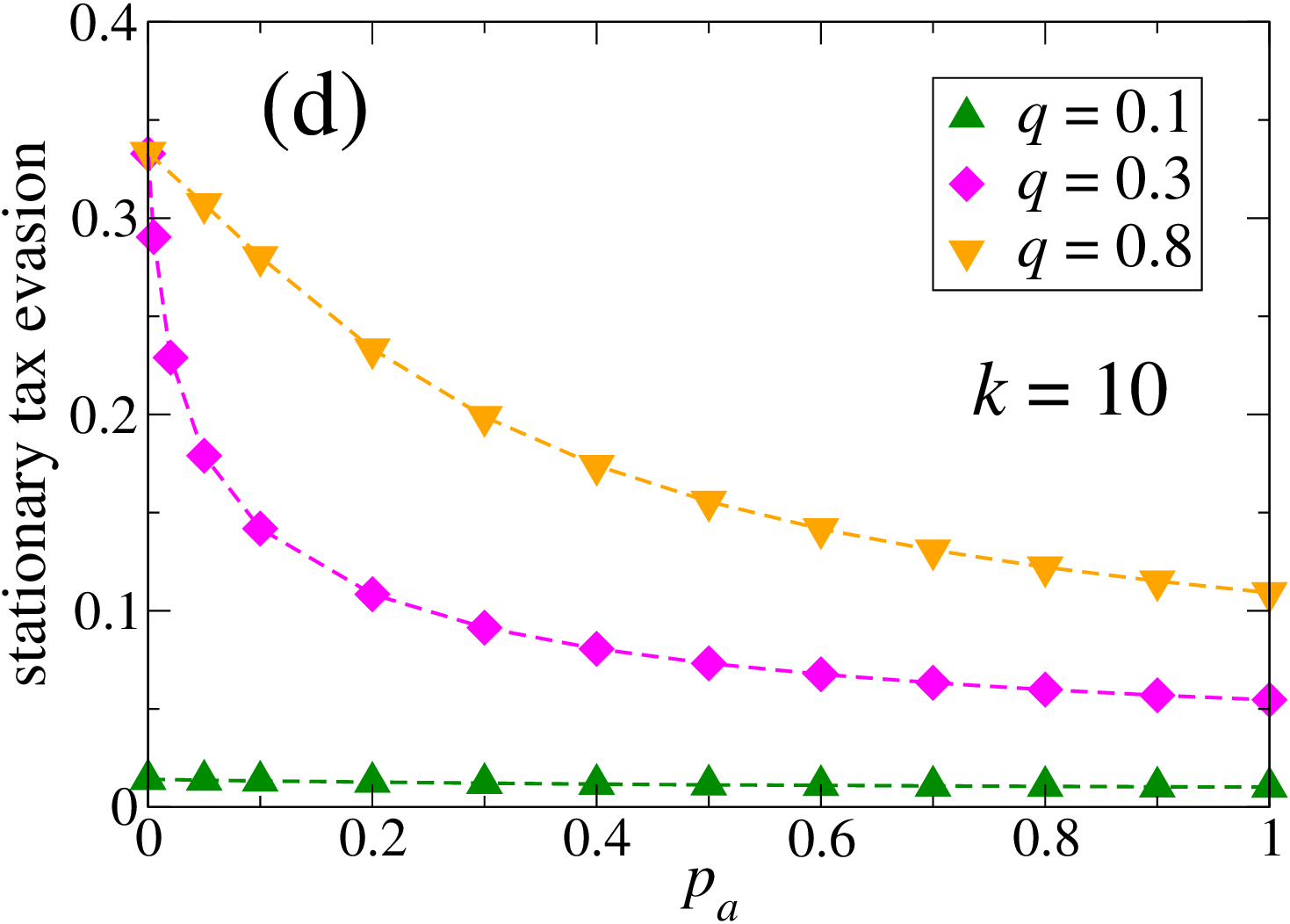}
\end{center}
\caption{(Color online) Average stationary tax evasion as a function of the audit probability $p_{a}$. In the panels (a), (b) and (c) the results are for $q=0.1$, $q=0.3$ and $q=0.8$, respectively, and typical values of $k$. In the last panel (d) it is shown the results for $k=10$ and different fractions of negative interactions $q$. Each point is averaged over $100$ independent simulations for population size $N=10^{4}$, and the dashed lines are just guides to the eye.}
\label{fig5}
\end{figure}

To better analyze the results obtained from the simulations, we exhibit in Fig. \ref{fig5} the average tax evasion in the stationary states, i.e., in the long-time limit, as a function of the audit probability $p_{a}$. In Figs. \ref{fig5} (a), (b) and (c) we present the results for $q=0.1$, $q=0.3$ and $q=0.8$, respectively, for typical values of $k$. In these panels one can clearly see the above-discussed effects. Indeed, for $q<q_{c}$ the tax evasion is low, but the punishment can effectively reduce such evasion if we consider large values of $k$ as $k=100$. On the other hand, the long-time tax evasion for the cases $q>q_{c}$ can be considerably decreased by the application of public policies. In other words, the tax evasion decreases for $\approx 33\%$ in the absence of punishment ($p_{a}=0$) until very small percentages like $\approx 1\%$. One can also see in these cases ($q>q_{c}$) that for large $k$ and $p_{a} > 50\%$ the compliance does not change considerably, suggesting that it is more important to monitor individuals caught in audits for a very long time than to perform extremely efficient audits. Similar conclusions can be obtained if one analyze the effect of changing $q$ for a fixed value of $k$, as shown in Fig. \ref{fig5} (d).

\begin{figure}[t]
\begin{center}
\vspace{0.5cm}
\includegraphics[width=0.48\textwidth,angle=0]{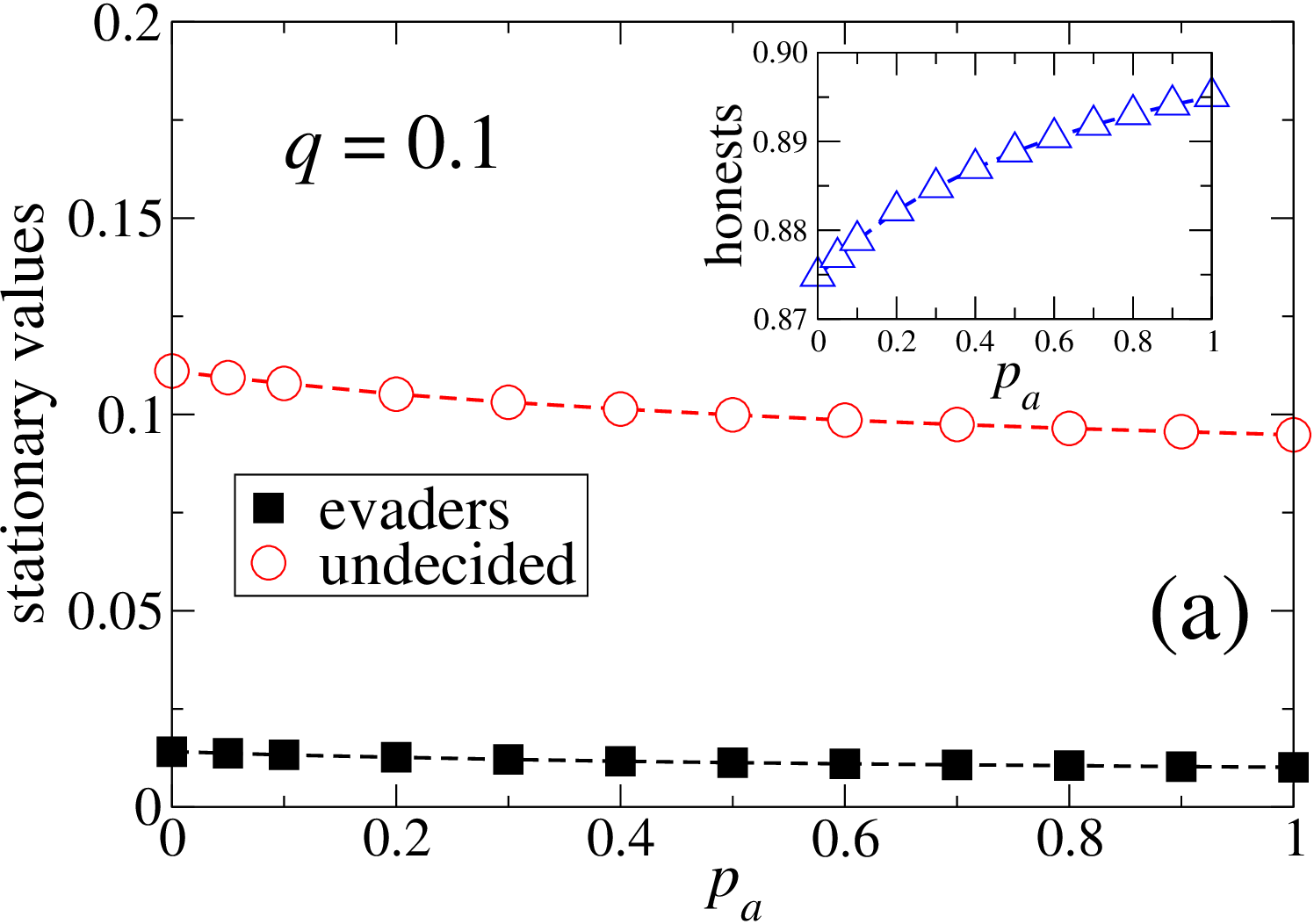}
\hspace{0.4cm}
\includegraphics[width=0.47\textwidth,angle=0]{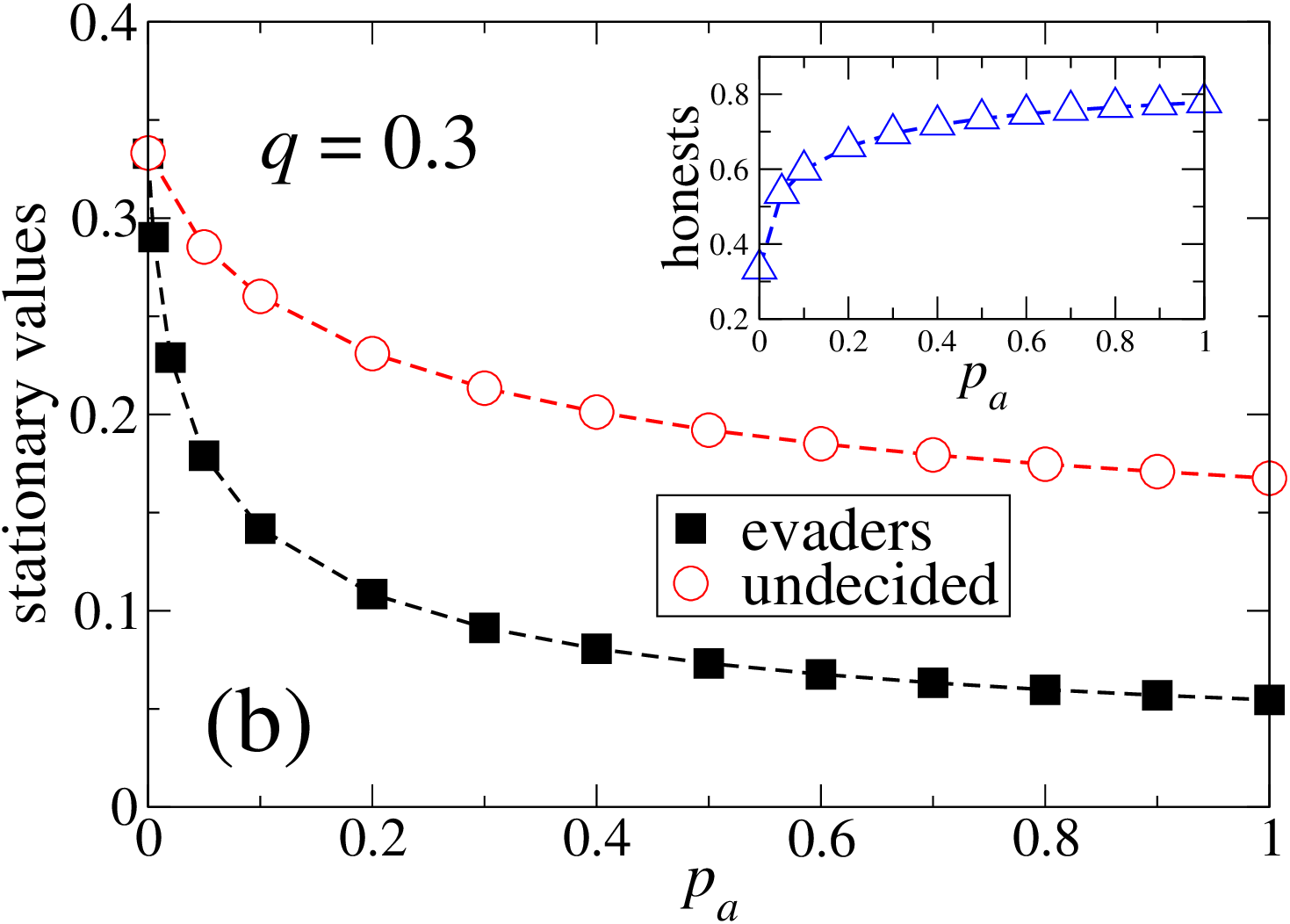}
\\
\vspace{0.90cm}
\includegraphics[width=0.47\textwidth,angle=0]{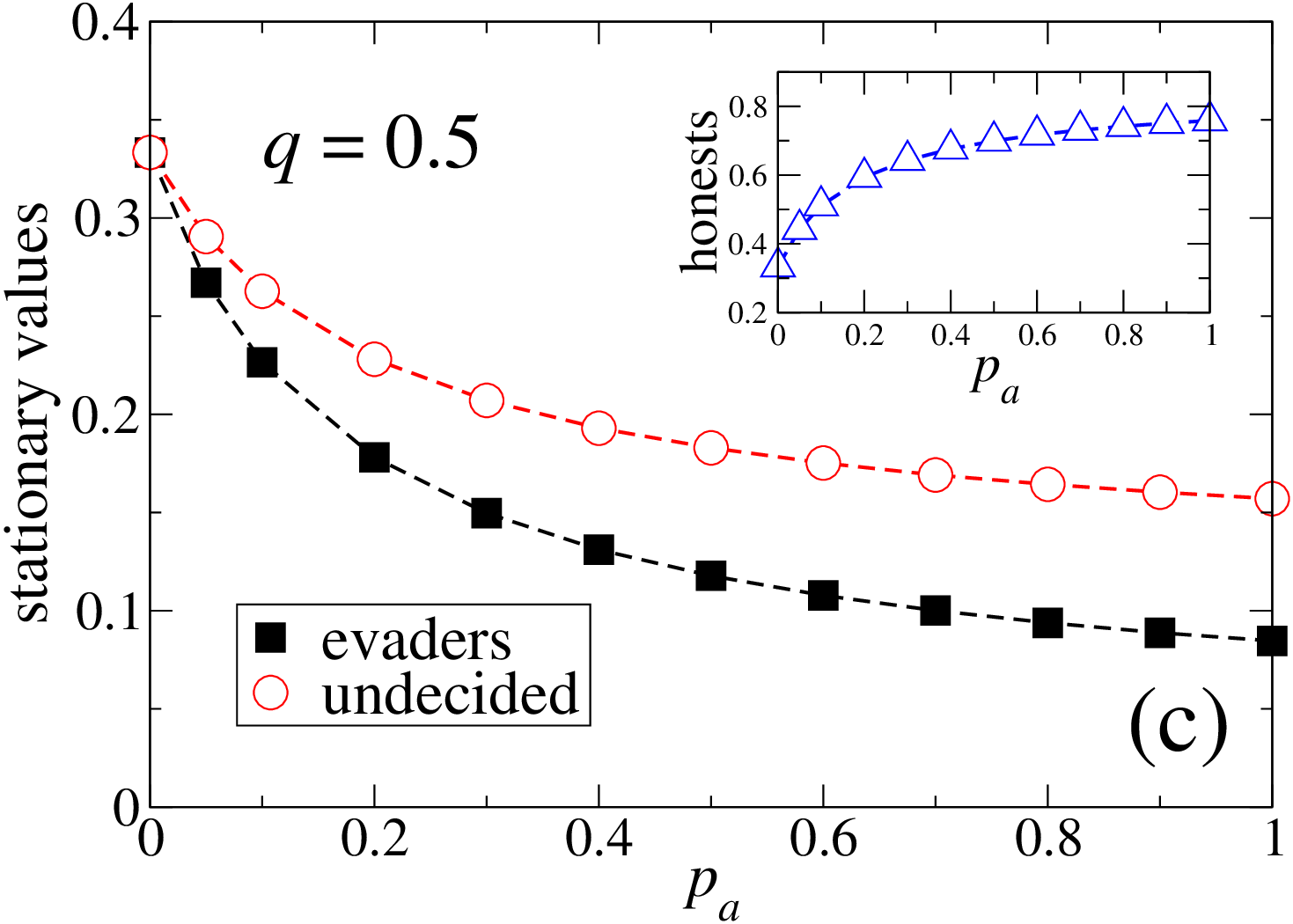}
\hspace{0.4cm}
\includegraphics[width=0.47\textwidth,angle=0]{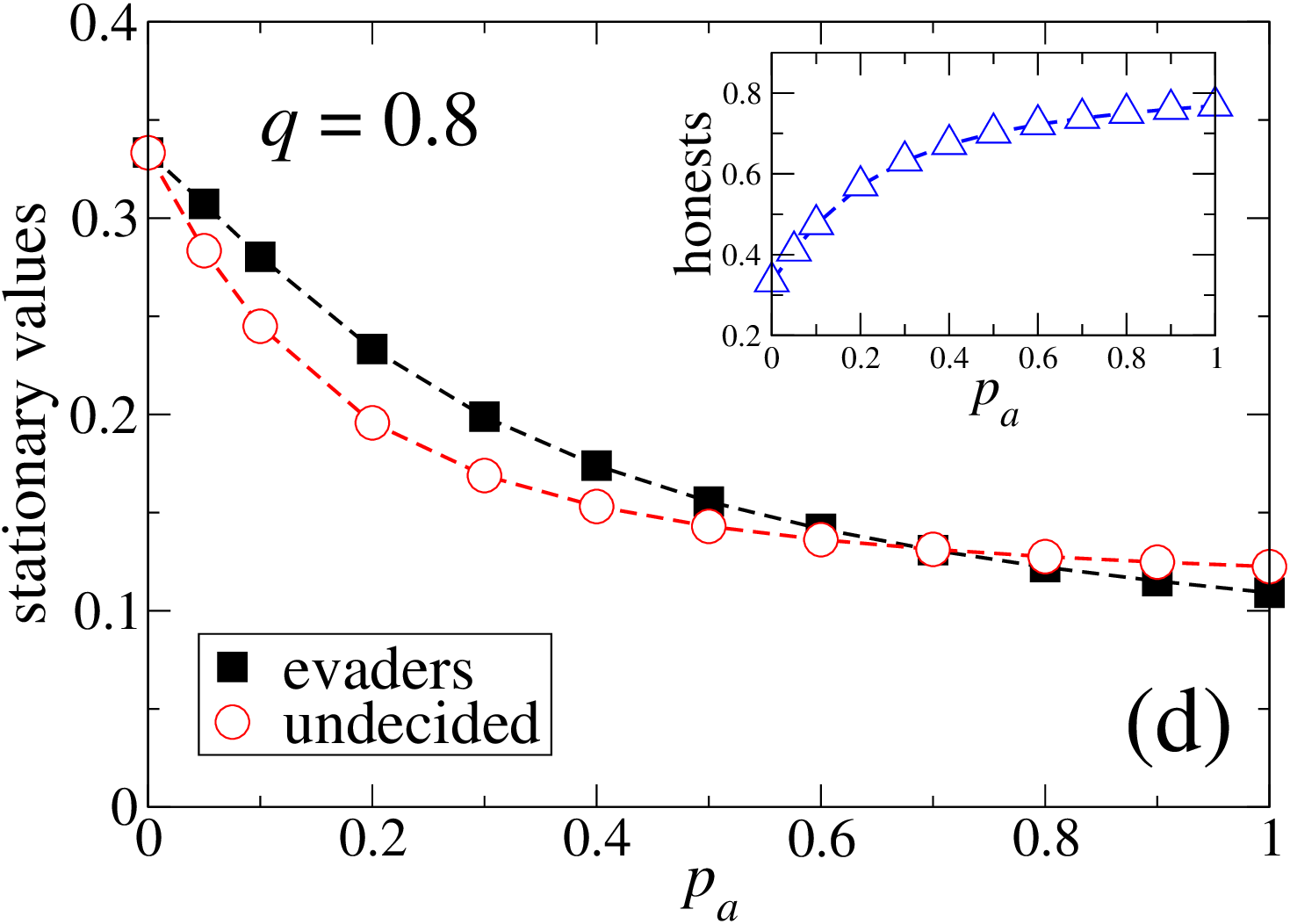}
\end{center}
\caption{(Color online) Average stationary fractions of evaders (squares), undecided (circles) and honests (triangles, in the insets) as functions of $p_{a}$ for $k=10$ and typical values of $q$, namely $q=0.1$ (a), $q=0.3$ (b), $q=0.5$ (c) and $q=0.8$ (d). Each point is averaged over 100 independent simulations for population size $N=10^{4}$, and the dashed lines are just guides to the eye.}
\label{fig6}
\end{figure}

After the analysis of the tax evaders' behavior, a question may arise: what is the behavior of the other two classes ($s=+1$ and $s=0$)? In particular, remembering that we are considering a 3-state model instead of an usual 2-state one, what is the impact of the presence of the third class, the undecided individuals, in the evolution of the system? In Fig. \ref{fig6} we exhibit the stationary fractions of the three classes of agents as functions of the audit probability $p_{a}$ for $k=10$ and typical values of $q$. One can see that, in general, the stationary fraction of undecided individuals is greater than the stationary fraction of tax evaders, and the former is always $> 10\%$ of the population. The evaders in the long-time limit are majority in comparison with undecided agents only for large densities of negative interactions $q$, like $q=0.8$, and $p_{a}<0.7$ [see Fig. \ref{fig6} (d)]. Even in this case, the honests are the majority in the population. For increasing values of $p_{a}$, the fraction of honest agents grows slowly for $q<q_{c}$ and fastly for $q>q_{c}$. One can also see from Fig. \ref{fig6} that when we rise the fraction of negative interactions $q$, it is more difficult to control the tax evasion in the population, as discussed above. Indeed, for $q=0.3$ we have a density of $\approx 10\%$ of tax evaders for $p_{a}=0.2$, a value that is only reached for $q=0.5$ and $q=0.8$ at $p_{a}\approx 0.7$ and $p_{a}\to 1.0$, respectively. This is due to the high disagreement among the individuals in the population. 

Summarizing, one can see that the presence of undecided agents in the population (state $s=0$) naturally reduces the number of evaders, and the implementation of public policies for punishment can be effective in controlling the tax evasion in the population, either due to fear of the undecided to being caught in an audit (if he/she become an evader), as by the monitoring of the tax evaders. This kind of behavior was not observed in the previous studies on the Zaklan model \cite{zaklan,lima1,lima2,lima3,seibold}.


\section{Final remarks}   

\qquad In this work, we have studied the dynamics of tax evasion on a fully-connected population. Different from the previous studies on the Zaklan econophysics model, where the agents can be in two distinct states ($s=\pm 1$), the dynamics of interactions among the agents in our model follows a three-state ($s=+1, -1, 0$) kinetic exchange opinion model \cite{biswas}, where individuals interact by pairs with competitive negative (with probability $q$) and positive (with probability $1-q$) couplings. Furthermore, we have considered the enforcement rules of the Zaklan model, where each agent is caught by an audit with probability $p_{a}$ and he/she is punished and remains honest during the following $k$ periods of time (or time steps).

Below the critical point $q_{c}=1/4$ of the opinion model, the dynamics leads the population to a state where the majority of the agents are honests ($s=+1$). In this case, the punishment rules do not affect considerably the system. On the other hand, for $q>q_{c}$ the kinetic exchange opinion model conducts the system to a disordered state where the three opinions coexist with equal fractions ($1/3$ for each one). In this case, we verified that the application of the enforcement mechanism can considerably reduce the tax evasion in the population. This reduction increases for decreasing values of $q$, which means that it is more difficult to control the compliance in populations with more disagreement among the individuals, i.e., with a large value of $q$.

We have also verified that the fraction of undecided or undecided individuals, i.e., agents with state $s=0$, affects the evolution of the tax evasion. These agents survive in the population in the long-time limit, which favors the reduction of the tax evasion. This fact together with the control given by the public policies may lead to low levels of evasion.

Regarding the critical behavior of the model, that is clearly observed in the absence of the enforcement rules (i.e., for $p_{a}=k=0$), one can easily see from our numerical results that the application of the punishment rules induces a decrease of the number of tax evaders ($s=-1$) in the system, and consequently an increase of the number of honest agents (state $s=+1$), which makes the order parameter defined in Eq. (\ref{eq3}) always greater than zero, i.e., we have no disordered phase in the presence of the enforcement rules (i.e., for $p_{a}>0$ and $k>0$).

As a future work, it can be interesting to analyze how different initial fractions of undecided individuals affect the tax evasion dynamics. In addition, the effects of agents convictions in the model can also be a realistic feature to be addressed. This kind of heterogeneity can affect considerably the evolution of opinion models \cite{jstat_pmco,celia,meu,conference,xiong}, and will certainly affect the dynamics of tax evasion considered here. Finally, the presence of inflexible agents \cite{galam} may also be important for better understanding of collective decision-making phenomena as occur in the dynamics of tax evasion.

\section*{Acknowledgments}

The author acknowledges financial support from Proppi - Universidade Federal Fluminense, Brazil, through the FOPESQ project.


\begin{thebibliography}{40}

\bibitem{galam_book}
S. Galam, \textit{Sociophysics: A Physicist's Modeling of Psycho-political Phenomena} (Springer, Berlin, 2012).

\bibitem{sen_book}
P. Sen, B. K. Chakrabarti, \textit{Sociophysics: an introduction} (Oxford University Press, Oxford, 2013).

\bibitem{econ_book}
\textit{Econophysics and Sociophysics: Trends and Perspectives}, edited by B. K. Chakrabarti, A. Chakraborti and A. Chatterjee (Wiley-VCH, Berlin, 2006).

\bibitem{pmco_book}
D. Stauffer, S. Moss de Oliveira, P. M. C. de Oliveira, J. S. S\'a Martins, \textit{Biology, Sociology, Geology by Computational Physicists} (Elsevier, Amsterdam, 2006).

\bibitem{bloom}
K. M. Bloomquist, Soc. Sci. Comput. Rev. 24, 411 (2006).

\bibitem{prinz}
M. Pickhardt, A. Prinz, J. Econ. Psychol. 40, 1 (2013).
%

\bibitem{andreoni} 
J. Andreoni, B. Erard, J. Feinstein, J. of Economic Literature 36, 818 (1998).

\bibitem{gachter}
S. Gachter, “Moral Judgments in Social Dilemmas: How Bad is Free Riding?” Discussion Papers 2006-03 CeDEx, University of Nottingham, 2006.

\bibitem{frey}
B. S. Frey, B. Togler, “Managing Motivation, Organization and Governance,” IEW-Working Papers 286, Institute for Empirical Research in Economics, University
of Zurich, 2006.


\bibitem{follmer}
H. Follmer, J. Math. Econ. 1, 51 (1974).

\bibitem{slemrod}
J. Slemrod, J. Econ. Perspect. 31, 25 (2007).

\bibitem{davis}
J. S. Davis, G. Hecht, J. D. Perkins, Account. Rev. 78, 39 (2003).

\bibitem{zaklan}
G. Zaklan, F. Westerhoff, D. Stauffer, J. Econ. Interact. Coord. 4, 1 (2009).

\bibitem{lima1}
G. Zaklan, F. W. S. Lima, F. Westerhoff, Physica A 387, 5857 (2008).

\bibitem{lima2}
F. W. S. Lima, G. Zaklan, Int. J. Mod. Phys. C 19, 1797 (2008).

\bibitem{lima3}
F. W. S. Lima, J. Theor. Economics Lett. 2, 87 (2012).


\bibitem{llacer}
T. Llaccer, F. J. Miguel, J. A. Nogueira, E. Tapia, Adv. Complex Syst. 16, 1350007 (2013).

\bibitem{seibold}
G. Seibold, M. Pickhardt, Physica A 392, 2079 (2013).
%

\bibitem{maj_vot}
M. J. de Oliveira, J. Stat. Phys. 66, 273 (1992).

\bibitem{biswas}
S. Biswas, A. Chatterjee, P. Sen, Physica A 391, 3257 (2012).

\bibitem{jstat_pmco}
N. Crokidakis, P. M. C. de Oliveira, J. Stat. Mech. P11004 (2011).

\bibitem{celia}
N. Crokidakis, C. Anteneodo, Phys. Rev. E 86, 061127 (2012).

\bibitem{meu}
N. Crokidakis, J. Stat. Mech. P07008 (2013).

\bibitem{conference}
S. Biswas, A. K. Chandra, A. Chatterjee, B. K. Chakrabarti, J. Phys.: Conf. Ser. 297, 012004 (2011).

\bibitem{xiong}
F. Xiong, Y. Liu, J. Zhu, Entropy 15, 5292 (2013).

\bibitem{galam}
S. Galam, F. Jacobs, Physica A 381, 366 (2007).




\end{thebibliography}
\end{document}